\documentclass[prb,aps,twocolumn,showpacs,superscriptaddress,floatfix]{revtex4}
\usepackage{epsfig}
\usepackage{epstopdf}
\usepackage{amsmath}
\usepackage{amssymb}
\usepackage{amsfonts}
\usepackage{mathptmx}
\usepackage{eucal}
\usepackage{bm}

\DeclareMathOperator{\re}{\mathop{\mathrm{Re}}}

\graphicspath{{Pictures1/}}

\begin{document}

\title{Protected 0-$\pi $ states in SIsFS junctions for Josephson memory and logic}
\author{S.~V.~Bakurskiy}
\affiliation{Skobeltsyn Institute of Nuclear Physics, Lomonosov Moscow State University
1(2), Leninskie gory, Moscow 119234, Russian Federation}
\affiliation{Moscow Institute of Physics and Technology, Dolgoprudny, Moscow Region,
141700, Russian Federation}
\author{N.~V.~Klenov}
\affiliation{Faculty of Physics, M.V. Lomonosov Moscow State University, 119992 Leninskie
Gory, Moscow, Russia}
\affiliation{Skobeltsyn Institute of Nuclear Physics, Lomonosov Moscow State University
1(2), Leninskie gory, Moscow 119234, Russian Federation}
\affiliation{Moscow Institute of Physics and Technology, Dolgoprudny, Moscow Region,
141700, Russian Federation}
\affiliation{All-Russian Research Institute of Automatics n.a. N.L. Dukhov (VNIIA),
127055, Moscow, Russia}
\author{I.~I.~Soloviev}
\affiliation{Skobeltsyn Institute of Nuclear Physics, Lomonosov Moscow State University
1(2), Leninskie gory, Moscow 119234, Russian Federation}
\affiliation{Moscow Institute of Physics and Technology, Dolgoprudny, Moscow Region,
141700, Russian Federation}
\author{N.~G.~Pugach}
\affiliation{Skobeltsyn Institute of Nuclear Physics, Lomonosov Moscow State University
1(2), Leninskie gory, Moscow 119234, Russian Federation}
\affiliation{National Research University Higher School of Economics, 101000, Moscow,
Russia}
\author{M.~Yu.~Kupriyanov}
\affiliation{Skobeltsyn Institute of Nuclear Physics, Lomonosov Moscow State University
1(2), Leninskie gory, Moscow 119234, Russian Federation}
\affiliation{Moscow Institute of Physics and Technology, Dolgoprudny, Moscow Region,
141700, Russian Federation}
\author{A.~A.~Golubov}
\affiliation{Moscow Institute of Physics and Technology, Dolgoprudny, Moscow Region,
141700, Russian Federation}
\affiliation{Faculty of Science and Technology and MESA+ Institute for Nanotechnology,
University of Twente, 7500 AE Enschede, The Netherlands}
\date{\today }

\begin{abstract}
We study the peculiarities in current-phase relations (CPR) of the SIsFS junction in the region of $0$ to $\pi $ transition. 
These CPR consist of two independent branches corresponding to $0-$ and $\pi-$ states of the contact. 
We have found that depending on the transparency of the SIs tunnel barrier the decrease of the s-layer thickness leads to transformation of the CPR shape going in the two possible ways: either one of the branches 
exists only in discrete intervals of the phase difference $\varphi$ or both branches are sinusoidal but differ in the magnitude of their critical currents. 
We demonstrate that the difference can be as large as $10\%$ under maintaining superconductivity in the s layer. An applicability of these phenomena for memory and logic application is discussed.
\end{abstract}

\pacs{74.45.+c, 74.50.+r, 74.78.Fk, 85.25.Cp}
\maketitle

Josephson junctions with ferromagnetic (F) layers in weak link region are
considered as promising control elements in a superconducting memory
compatible with RSFQ logic circuits \cite{Eschrig1, Linder1, Blamire1, JMRAM61, Mukhanov1,soloviev1}.
The presence of two or more ferromagnetic layers in the weak-coupling region
makes it possible to control the magnitude of the critical current $J_{C}$
of these junctions by changing of mutual orientation of F films
magnetization vectors \cite{RevG1, RevB1, RevV1, Bell, Qader, Baek,  Birge2018}. It is necessary to
mention that the large number of ferromagnetic layers in the weak-coupling
area is accompanied by degradation of $J_{C}$ by virtue of the larger number
of interfaces in the structure, and owing to the strong suppression of
superconducting correlations in each of the F layers.

In \cite{Sobanin, Larkin, Vernik, SIsFSAPL, Caruso1, Caruso2} it was shown that the required changes
in $J_{C}$ can also be ensured in SFS or SIsFS structures with single
ferromagnetic layer. The remagnetization of the ferromagnetic layer
in these junctions shifts the position of the maximum in Fraunhofer-like dependence of $J_{C}$ on
external magnetic field resulting in changing of $J_{C}$ magnitude at zero
field. This principle was extended in magnetic rotary valves \cite{RotWal1, SF-NFSP} where the switching effect in $J_{C}$ magnitude was
achieved by changing the direction of the inplane F film magnetization.

It should be noted that magnetization reversal processes significantly
increase the characteristic response time of the SFS control memory elements
in comparison with the characteristic switching time of Josephson contacts
in SFQ  logic  circuits. In order to overcome this drawback, it was suggested
in \cite{PhaseDom} to use
SIs-F/N-S contacts, where thin s-layer can be subdivided on superconducting domains with a phase shift of $\pi$. However, the implementation of the above mentioned proposals is
a rather complicated technological task.

The promising concept of the Josephson memory with electrical control can be also realized using the phenomenon of the coexistense of the two metastable states 
in the vicinity of $0$ - $\pi $ transition \cite{GoldMemory, GoldButterfly}. For instance, these states can be achieved inside the region of the $0-\pi$ transition of the junction with ferromagnetic layer \cite{Klapwijk, Frolov, Baek2018, Frolov2018} or in the junctions with two uncollinearlly magnetized hard ferromagnets \cite{Trifunovic, Houzet2}. The conditions for
the existence of metastable states essentially depend both on the material
parameters of the contacts and on their geometry. 

The purpose of this article is to propose the concept of the control element for memory based on our finding of noticable diifference between critical current in $0$- and $\pi$- states in certain range of the s-layer thickness $d_s$
in SIsFS junctions.

\begin{figure}[h]
\begin{minipage}[h]{0.99\linewidth}
\center{a)\includegraphics[width=0.99\linewidth]{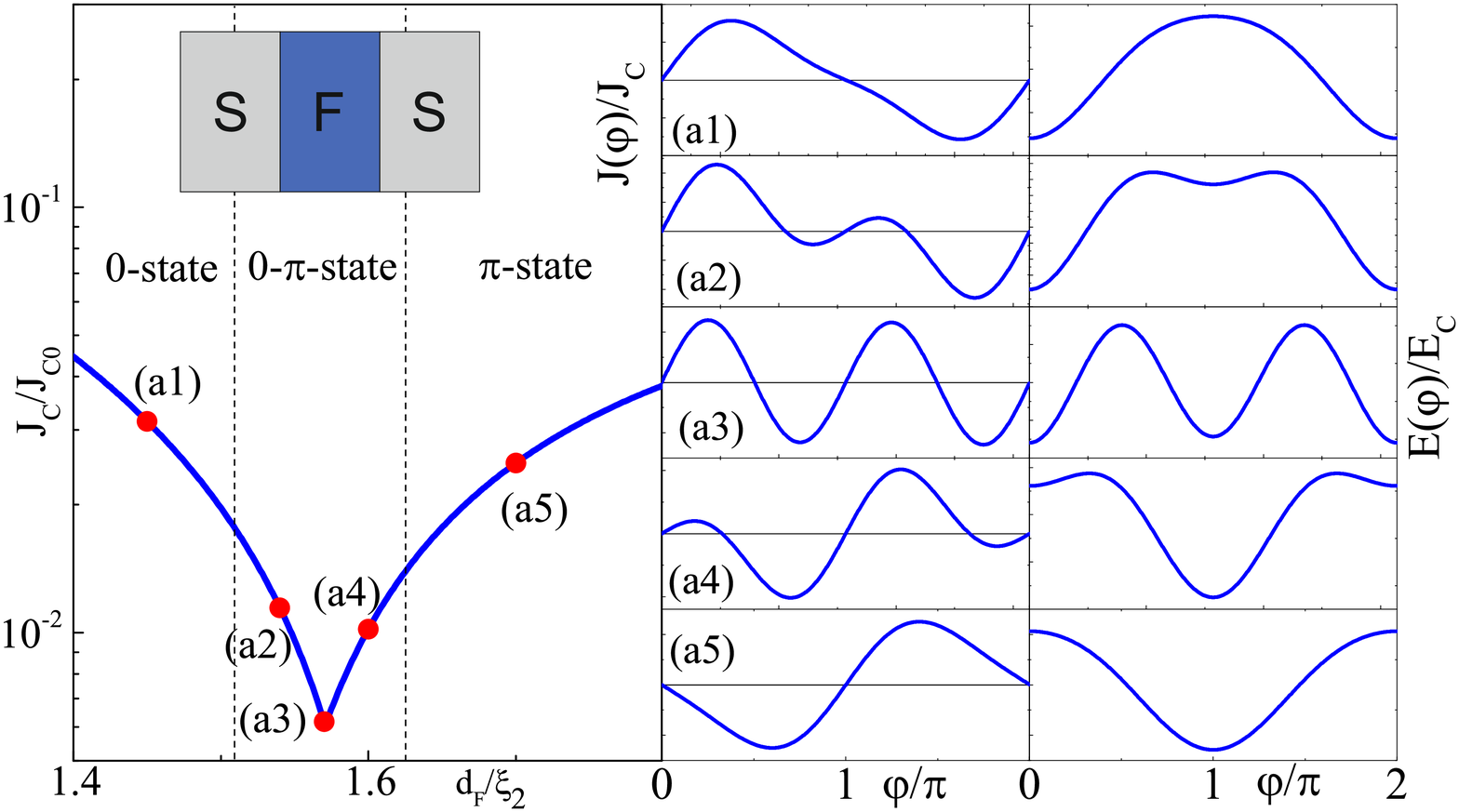}}
\vspace{-2 mm}
\end{minipage}
\par
\vfill
\begin{minipage}[h]{0.99\linewidth}
\center{b)\includegraphics[width=0.99\linewidth]{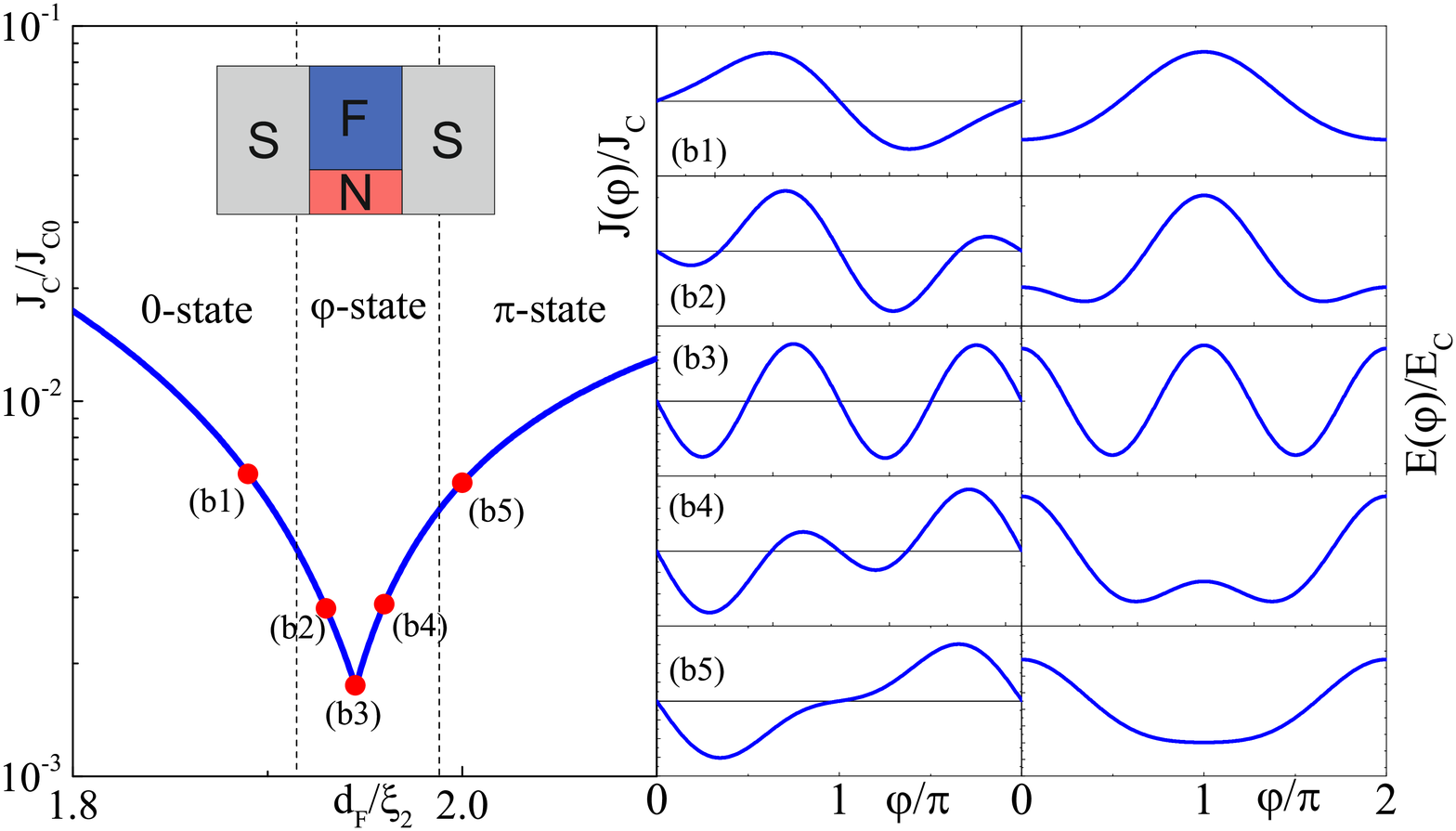}}
\vspace{-2 mm}
\end{minipage}
\vfill
\begin{minipage}[h]{0.99\linewidth}
\center{c)\includegraphics[width=0.99\linewidth]{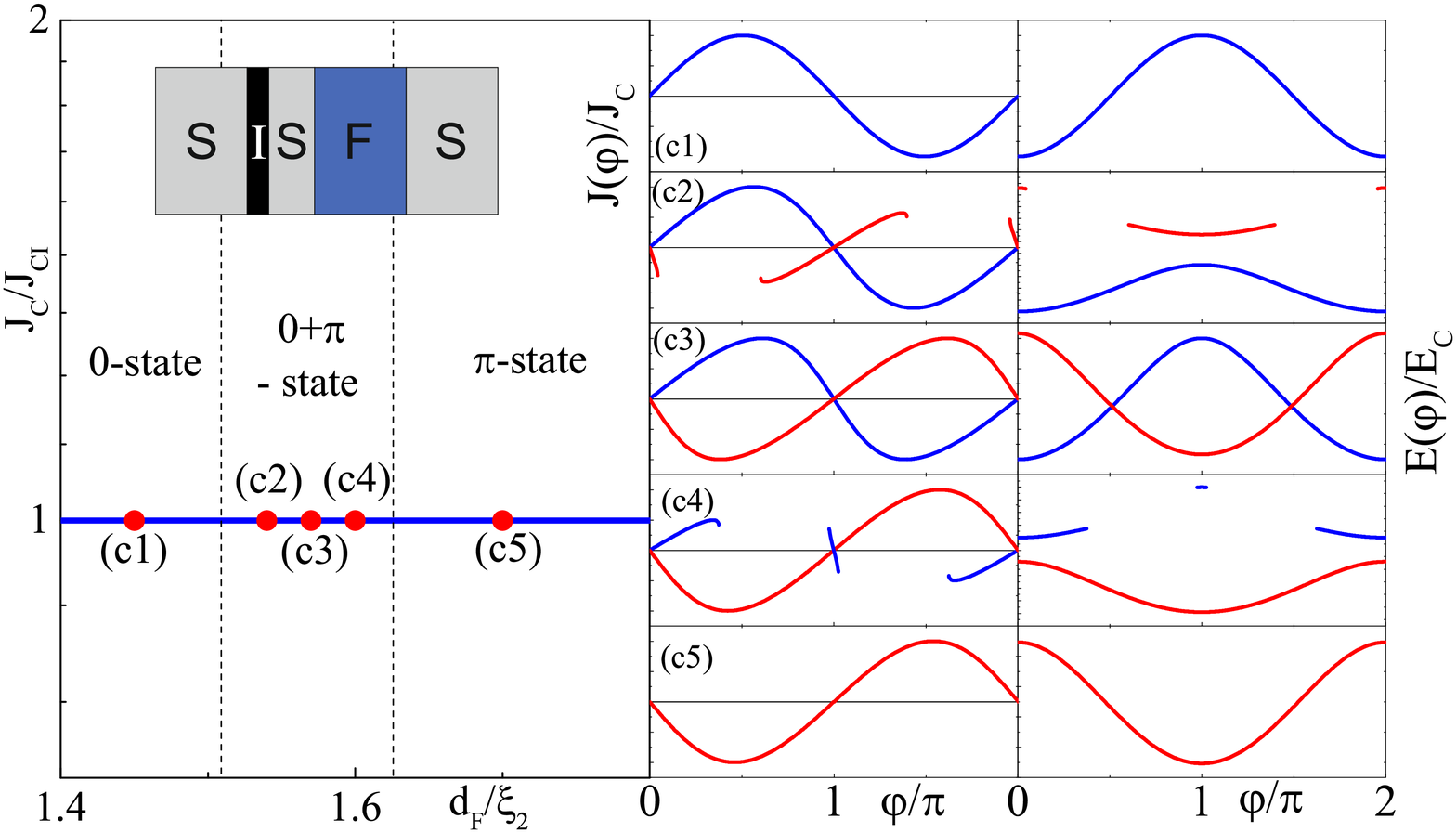}}
\vspace{-2 mm}
\end{minipage}
\caption{Typical behaviour of a) SFS, b) S-F/N-S and c) SIsFS junctions
characteristics in a vicinity of $0-\protect\pi$ transition. Each block a)-c) includes 3 parts. The left one
demonstrates dependence of the critical current $J_C$ versus the ferromagnet
layer thickness $d_F$. The middle row contains
CPRs of the junctions at the certain $d_F$ marked by the red dots (1)-(5)
on the $J_C(d_F)$ dependence. Finally, the right row demonstrates EPRs at
the same points.}
\label{Intro}
\end{figure}

Figure \ref{Intro} schematically shows the principle of operation of SIsFS structure in comparison with SFS or S-F/N-S devices. This figure demonstrates 
the evolution of the current-phase relation (CPR) $J_{S}(\varphi )$ and energy-phase relation (EPR) $%
E_{S}(\varphi )$ of SFS or S-F/N-S junctions and SIsFS structures in a
vicinity of $0$ to $\pi $ transition, which occurs with increase of F layer
thickness $d_{F}$.

For SFS junction the amplitude of the second harmonic in CPR is positive \cite%
{Buzdin2005, Buzdin1, Baek2018,  Frolov2018} at any transition point. The $0-\pi $
transition is going through region of coexistence of the states with $0$ and 
$\pi $ phase differences (See Fig.\ref{Intro}a). During the transition the
depths of the corresponding minima in the $E_{S}(\varphi )$ relation are changing continuously 
until one of them disappears.

In the Josephson junctions with parallel $0-$ and $\pi -$ channels for the
supercurrent flow inside a weak link region \cite{Buzdin, Gold, Pugach, Gold2, Bakurskiy3}, e.g. S-F/N-S contacts, the amplitudes of the first harmonic
in CPR in the channels have opposite signs and compensate each other. At the
same time, the amplitudes of the second harmonic have negative signs
in both channels. In this situation the transition from the initial $0$
state to the final $\pi $ state takes place via formation of so-called $\varphi $%
-state. With the increase of $d_{F}$ (see Fig.\ref{Intro}b) the minimum in $%
E_{S}(\varphi )$ located at $\varphi =0$ splits into two minima located at
some, $\pm\varphi (d_{F}),$ and this $\varphi (d_{F})$ tends to $\varphi =\pi $
at the end of the transition.

In both $0-\pi $ transitions presented in Fig.\ref{Intro}a,b the $%
J_{C}(d_{F})$ curves are V-shaped with a strong suppression of the critical
current during the transitions.  The metastable states in both structures generally have different barriers, which would be exceeded to switch device into resistive regime. It provides opportunity to read the state, although after that the state will be erased.

Contrary to that, in SIsFS\ devices \cite{SIsFS2013,Nevirkovets1,Ruppelt,Nevirkovets2, SIsFS2017} it is possible to
realize the mode of operation in which magnitude of $J_{C}$ of SIs part of
the structure is smaller than the amplitude of the second harmonic in CPR of
its sFS part. In this case (see Fig. 1c) $|J_{C}|$ is constant during the $%
0-\pi $ transition \cite{Ruppelt, SIsFS2017}, although the current-phase relations undergo
significant transformations and become multivalued. For relatively large layer thickness $d_s$ there
is the domain of SIsFS junction parameters providing its stay either in $0$-
or in $\pi $-ground state.  The critical current is determined by SIs part of the junction and is exactly the same for the both states.
In this domain a transition from one ground state to another is not possible
by a continuous adiabatic variation of the phase $\varphi $ and the states
can be used for storage an information since the energy barrier separated them is higher than the energy of tunnel junction, thus protecting the system
against accidental switching.

In order to model the SIsFS structure we suppose that the condition of a dirty limit is fulfilled for all
metals and that effective electron-phonon coupling constant is
zero in F layer. Under the above conditions the problem can be
analyzed in the framework of the Usadel equations \cite{Usadel} with
Kupriyanov-Lukichev boundary conditions \cite{KL} at the interfaces.
The boundary-value problem was solved numerically using
the algorithm developed in Ref. \cite{SIsFS2017}. For simplicity we assume below that the resistivities $\rho$ and coherence lengths $\xi=(D/2\pi
T_{C})^{1/2}$ of the SIsFS junction materials are the same. Here $T_{C}$ is a critical temperature and $D$ is a diffusion coefficient of superconducting material.

\begin{figure}
\center{\includegraphics[width=0.92\linewidth]{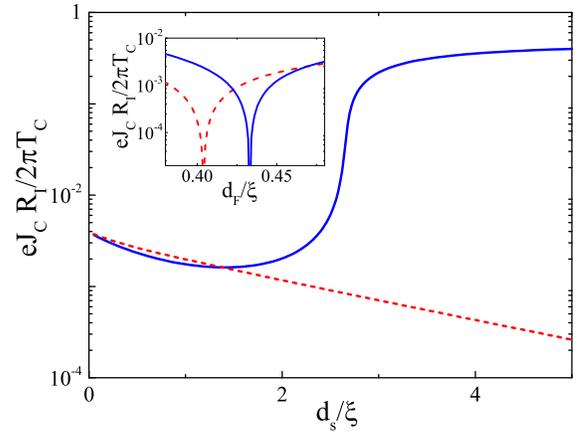}}
\vspace{-4 mm}
\caption{Magnitude of the critical current $J_C$ of the SIsFS ( solid blue) and the
SInFS (dashed red) junctions versus thickness of the middle layer $d_s$
calculated in the vicinity of $0-\protect\pi$ transition at $d_F=0.46\protect%
\xi$ and $T=0.26 T_C$. Inset: Critical current $J_C$ of the SIsFS (solid blue) and
the SInFS (dashed red) junctions versus thickness of the F-layer $d_F$ for s-layer
thickness $d_s=1\protect\xi$  much smaller than $d_{sC}$.}
\label{JCds}
\end{figure}

Based on our previous investigations \cite{SIsFS2017, SIsFS2013} we have fixed
the set of SIsFS\ junction parameters that ensure an occurrence
of SIsFS contact in the vicinity of the $0-\pi $ transition at large s layer
thickness $d_{s}=5\xi$: $d_{F}=0.46\xi $,  $T=0.26T_{C}$,  exchange energy $H=10\pi T_{C},$ suppression parameters of SF interface $%
\gamma _{BSF} = R_{BSF}\mathcal{A}/\xi \rho  = 0.3$ and SIs interface  $\gamma
_{BI}=R_{BI}\mathcal{A}/\xi \rho =5000$. Here $R_{B}$ and $\mathcal{A}$ are the resistance and area of corresponding interface. With this choice of parameters the weakest link of the SIsFS structure is located at the tunnel barrier thus providing the coexistence of the two
independent CPR branches (see example in the panel (c3) in Fig. \ref{Intro}c). 

We start with calculation of the dependence of $J_{C}$ magnitude
on $d_{s}$, shown as a blue line in Fig. \ref{JCds}. It has a common form with a rapid drop of the critical current near the critical thickness $d_{sC} \approx 2.7 \xi$. There are two independent processes 
going in the vicinity of this point. The first one is a shifting of the position and narrowing of the width of the $0-\pi$ transition during the decrease of the $d_{s}$, due to the change of the effecting boundary condtions on F layer \cite{Vernik}. Inset in the Fig. \ref{JCds} shows that transformation significantly differs from the process studied in the SInFS junction \cite{Pugach1}, since the residual pairing locks the phases of different Matsubara frequencies. This phenomenon leads to collapse of the 0-branch of the CPR at the $d_s$ near the critical point.

The second phenomenon is the deviation of the pair amplitude in the thin s-electrodes for $0$ and $\pi$ states due to the different symmetry of the anomalous Green functions predicted and found experimentally in the sFs junctions with thin s-electrodes \cite{Samokhvalov1, Samokhvalov2, Samokhvalov3}.
This effect can modify the critical currents of the tunnel SIs junction due to the changing of the s-layer properties in $0$ and $\pi$ states.

The relative impact of these processes on the SIsFS structure properties depends on the tunnel suppression parameter $\gamma_{BI}$. First, we consider the system at $\gamma_{BI}=5000$, when the collapse of the 0-branch occurs.



\begin{figure}[t]
\begin{minipage}[h]{0.99\linewidth}
\center{a)\includegraphics[width=0.95\linewidth]{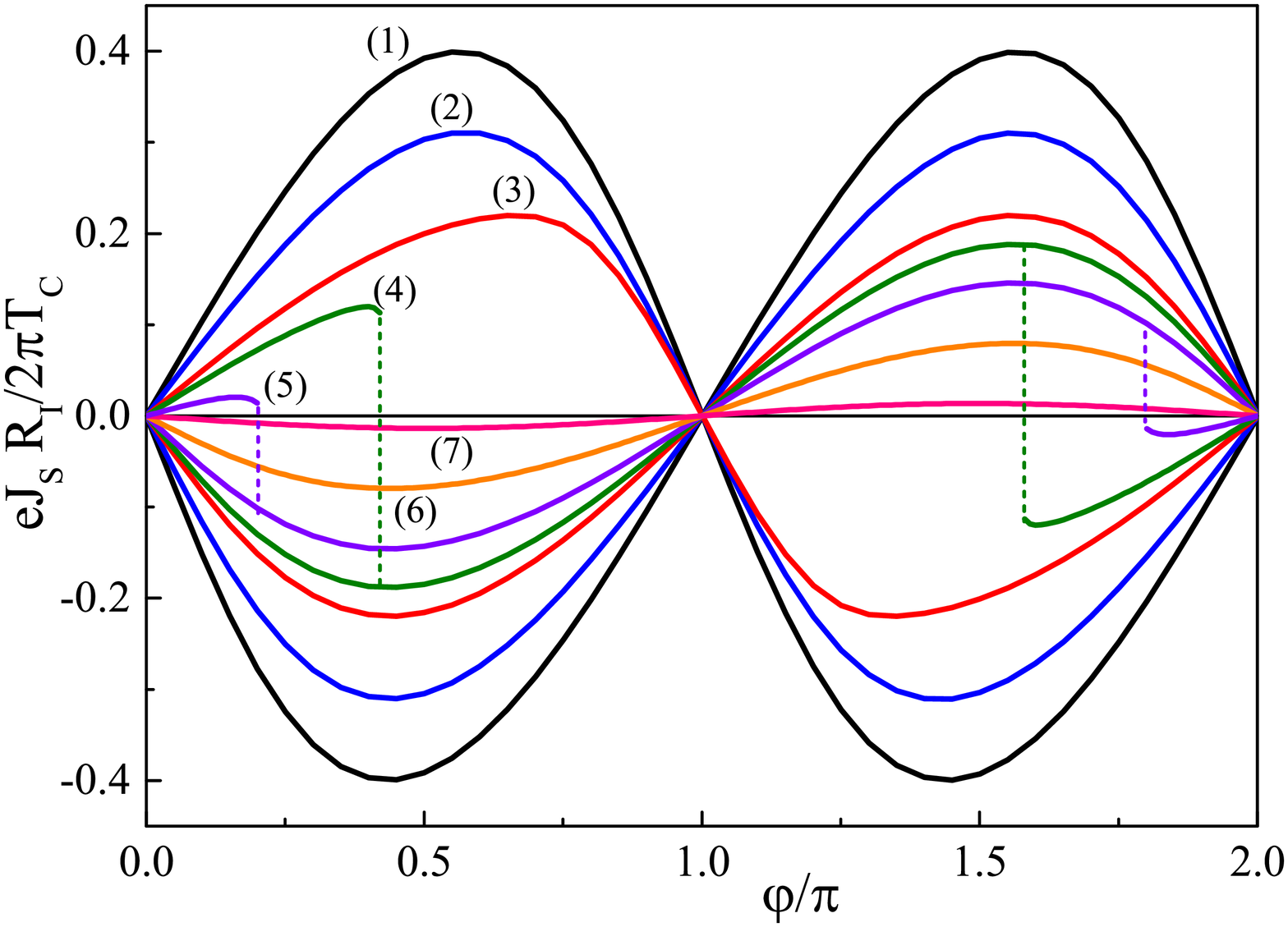}}
\end{minipage}
\par
\vfill
\vspace{-2 mm}
\begin{minipage}[h]{0.99\linewidth}
\center{b)\includegraphics[width=0.95\linewidth]{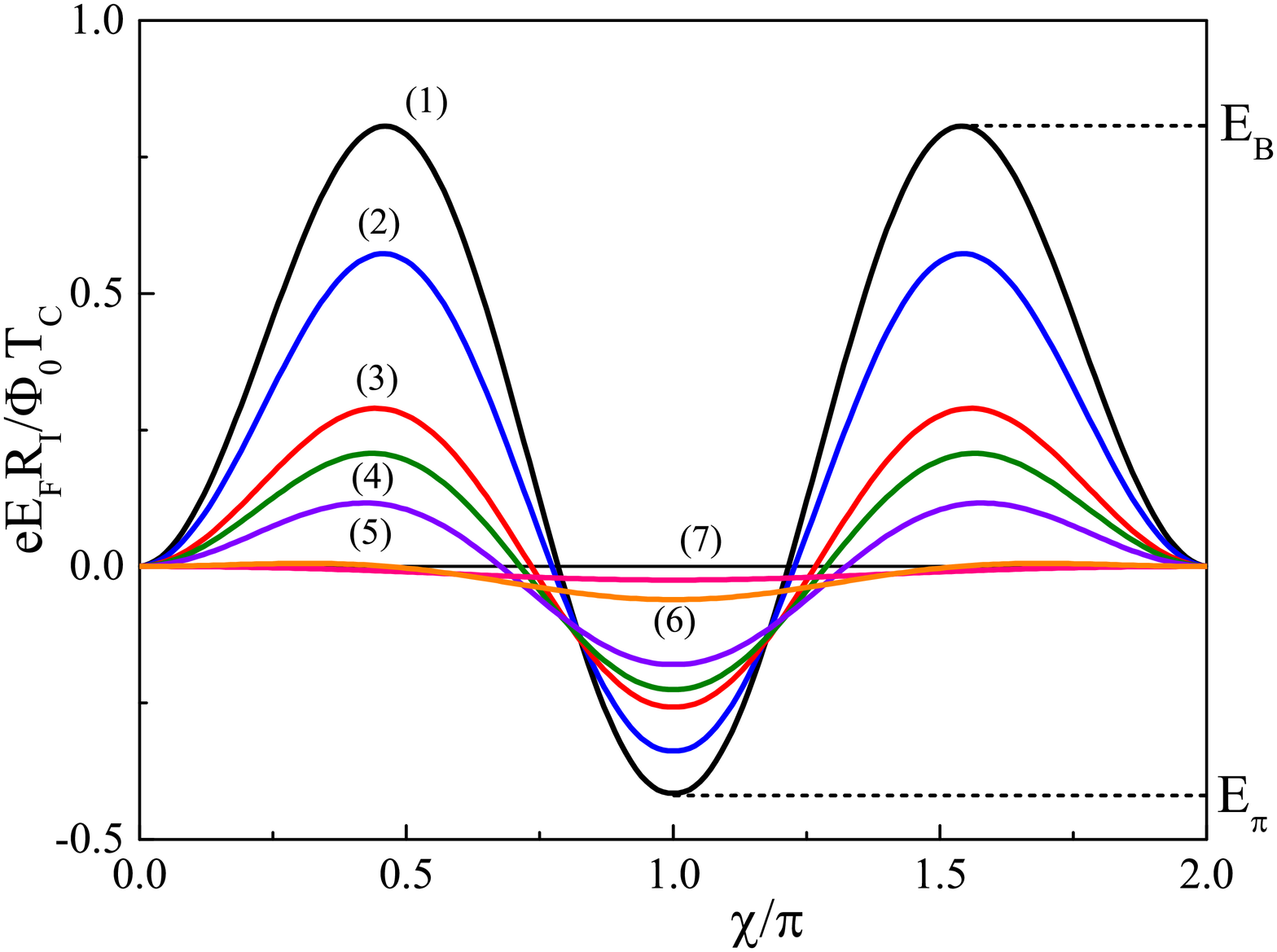}}
\end{minipage}
\vspace{-2 mm}
\caption{The current-phase relation $J_S(\protect\varphi)$ of the SIsFS
junction (a) and the energy-phase relation $E_F(\protect\chi)$ of the
corresponding sFS junction (b) for the set of the different thicknesses of
the s-layer $d_{s}$: (1) $5 \protect\xi$, (2) $3.5 \protect\xi$, (3) $3 
\protect\xi$, (4) $2.9 \protect\xi$, (5) $2.8 \protect\xi$, (6) $2.7 \protect%
\xi$, (7) $2.6 \protect\xi$. CPRs include $0$ and $%
\protect\pi$ branches in the cases (1)-(5) and only $\protect\pi$ branch for
(6)-(7). The calculations were done in the vicinity of $0-\protect\pi $
transition at $d_{F}=0.46\protect\xi $, $T=0.26T_{C}$ and $\protect\gamma %
_{B}=5000$.}
\label{CPRvsds}
\end{figure}

Figure \ref{CPRvsds}a demonstrates the evolution of the shape of $%
J_{S}(\varphi )$ dependence with decrease of the s layer thickness. It is
seen that the magnitudes of critical current of the both CPR\ branches
are monothonically decreasing, while the shape of the curves transforms in
different ways. The CPR\ of the $\pi $ ground state remains sinusoidal at
any $d_{s}$ during the $d_{s}$ decrease. The CPR
at the some $d_{S}$ the $0$-branch becomes broken and
disappears in the significant interval of phases. At $d_{s} \lesssim 2.7\xi $
it completely disappears and only the $\pi $ ground state still exists. At
the smaller $d_{s}\lesssim $ $2.5\xi $ critical current of the $\pi $-branch
is strongly suppressed due to intensive suppression of the superconductivity
in the s-layer by inverse proximity effect.

To understand the physics behind the $J_{S}(\varphi )$
transformations shown above it is convenient to use the so-called lump junction model 
\cite{SIsFS2017} and consider the SIsFS junction as a series connection of SIs
and sFS contacts with finite thickness $d_{s}$ of the s electrode. The self-consistent problem for sFS junction was solved numerically
with free boundary condition $d\Phi _{s}/dx=0$ at the Is interface. The choice of the initial pair potential in iterative self-consistent calculation permits to find the 
magnitudes of Usadel functions $\Phi _{s}$ at Is interface for both 0 and $\pi$-states.
 Taking
into account that the supercurrent flowing across the SIsFS structure is
essentially small compare to s and S films depairing current, and that the $%
d_{s}$ thicknesses of interest exceed $2.5\xi $ we can neglect a
dependence of $\Phi _{s}$ magnitudes on phase differences $\chi $ and $\chi
_{1}$ across both sFS and SIs junctions, respectively. Under this assumption
the critical current, $I_{C},$ and energy phase relation, $E_{I}(\chi _{1}),$
of SIs contact can be calculated using standard well-known expressions%
\begin{gather}
\frac{eI_{c}R_{I}}{2\pi T_{C}}=\frac{T}{T_{C}}\re\sum_{\omega >0}\frac{%
\Delta _{0}\Phi _{s}}{\sqrt{(\omega ^{2}+\Delta _{0}^{2})(\omega ^{2}+\Phi
_{s}^{2})}},  \label{Ictun} \\
E_{I}(\chi _{1})=E_{CI}(1-\cos (\chi _{1})),\quad E_{CI}=\frac{\Phi _{0}I_{c}%
}{2\pi },  \label{EhiTun}
\end{gather}%
where $R_{I}$ is the resistance of the tunnel barrier and $\Phi _{0}$ is the
magnetic flux quantum.

In the considered approximation the same quantities $\Phi _{s}$ can be used
as boundary conditions of the first kind in calculating CPR $J_{S}(\chi )$
and EPR%
\begin{equation}
E_F(\chi )=\frac{\Phi _{0}}{2\pi }\int_{0}^{\chi }J_{S}(\mu )d\mu ,
\label{EPRsFS}
\end{equation}%
of sFS contact (see Fig. \ref{CPRvsds}b).

Figure \ref{CPRvsds}b demonstrates that $E_{F}(\chi )$ dependence have the double-well form with the two minima at $\chi =\pi $ and $\chi =0$ separated by a potential
barrier $E_B$. 
The decrease of $d_{s}$ is accompanied by suppression of the barrier height, 
$E_{B}(d_{s}).$ At $d_{s}\lesssim 2.7\xi $ the potential barrier completely
disappears and the sFS contact stays only in the $\pi $ ground state. The depth of the potential well for the $\pi$-state
$E_\pi$ also decreases rapidly with further decrease of $d_{s}.$

\begin{figure}[t]
\center{\includegraphics[width=0.99\linewidth]{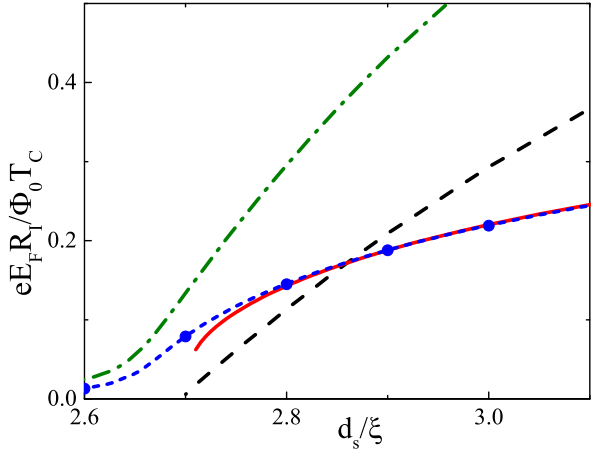}}
\vspace{-3 mm}
\caption{The characteristical energies of the SIs and sFS junctions: barrier heights in sFS junction $E_B$ (dashed black) and $E_{B}-E_{\pi}$ (dash-dotted green) and the energy of SIs junction $E_{CI}$ in 
the 0- (solid red) and $\pi$- (short-dashed blue) states versus the thickness of s layer $d_s$. The blue dots mark the energies calculated from the CPR of SIsFS junction demonstrated on Fig. \ref{CPRvsds}a.} 
\label{EB}
\end{figure}

We summarize the dependence of the characteristical energies $E_B$ (dashed black), $E_B-E_\pi$ (dash-dotted green) and $E_{CI}$ versus thickness $d_s$ in Fig. \ref{EB}. The energy of tunnel junction $E_{CI}$ is calculated in the frame of lumped junction model independently for 0- (solid red) and $\pi$- (short-dashed blue) states of sFS-electrode. It is seen, that they practically coincide with each other for $d_s > 2.8 \xi$. The blue dots mark the critical energies of the SIsFS junction shown in the Fig. \ref{CPRvsds}. They correllate well with the results obtained in lumped junction model thus demonstrating the accuracy of our approach.

It follows from Fig. \ref{EB}, that in the wide range of $d_s$ the tunnel energy $E_{CI} \ll E_{B}, E_{B} - E_{\pi}$. These conditions provide existence of two independent continuous branches of CPR in SIsFS junction 
for the both $0$- and $\pi$-states as it is seen in Fig. \ref{CPRvsds}a. With the decrease of $d_{s}$, the barrier height 
$E_{B}$ is suppressed more rapidly than $E_{CI}$. At $d_{s}\approx 2.8\xi $ the energies $%
E_{B}$ and $E_{CI}$ become comparable and adiabatic
increase of $\varphi $ leads to escape of the system from the metastable to
the ground state resulting in discontinues jump in $J_{S}(\varphi ).$ Finally, at the critical thickness $d_{s}\approx 2.7\xi $ the barrier completely vanishes, 
leading to dissappearence of the 0-branch. At the same time, the barrier for the $\pi$-state significantly exceeds $E_{CI}$ in the whole considered interval.

In the region near the critical thickness $d_s \approx 2.7\xi$ the energies of tunnel SIs junction corresponding to $0$- (red) and $\pi$- (blue) states deviate. 
The absolute values of Green function $\Phi_s$ and pair potential $\Delta$ on the tunnel SIs interface are different for 0 and $\pi$ states. In contrast to Ref.\cite{Samokhvalov1, Samokhvalov2, Samokhvalov3}, we find that $\pi$-state has a larger value of $\Delta$ and larger critical current.

The difference between $0$ and $\pi$-states provides significant influence on the CPR and critical current of SIsFS structure at the larger values of tunnel layer parameter $\gamma_{BI}$. Increase 
of $\gamma_{BI}$ doesn't modify the barrier height $E_B(d_s)$ of the SFs-junction, but proportionally decreases the $E_{CI}(d_s)$, shifting the cross-point between them to the critical thickness $d_{sC}$. 

\begin{figure}[t]
\center{\includegraphics[width=0.99\linewidth]{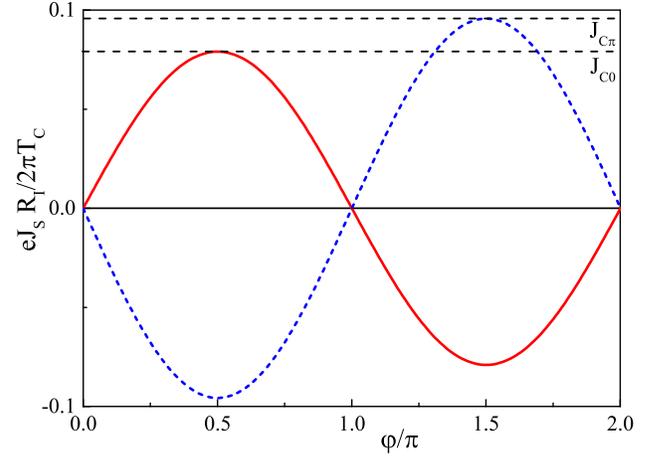}}
\vspace{-3 mm}
\caption{a) The $0-$ (solid red) and $\pi-$ (dashed blue) branches of the CPR $J_S(\protect\varphi)$ of the SIsFS
junction with thin s-layer $d_s =2.72 \xi$ and strong $\gamma_B = 5\cdot 10^7$.}
\label{CPRFinal}
\end{figure}

For instance, for the value of suppression parameter $\gamma_{BI} = 5\cdot10^7$ direct calculations of SIsFS structure show that the both branches of CPR have sinusoidal shape and are defined for all phases $\varphi$  even at $d_s=2.72 \xi$ (See Fig. \ref{CPRFinal}). It is important to note that for this particular case the critical currents $J_{C0}$ and $ J_ {C\pi}$ of $ 0$- and $ \pi$- CPR branches are different from each other and this difference $ (J_ {C \pi} -J_ {C0}) / J_ {C0} $ is of the order $ 10 \% $.

The ability of the structure to be in one of two states differing in their critical currents can find applications in superconductor logic and memory devices. Information on the state of the SIsFS structure can be obtained by setting a current pulse, with an amplitude, $J$ in the range $ J_ {C0}<J< J_ {C \pi}$. It is determined by the absence ($ \pi- $ state) or by the generation ($0- $ state) of the corresponding voltage pulse.  Reading the state of the element will be non-destructive, if the incoming energy is insufficient to open the channel of tunneling through the energy barrier separating the $0-$ and $\pi-$ states. The recording is possible with current pulses, which have an amplitude exceeding the critical current of sFS junction and switch the system from 0 to $\pi$- state and back \cite{GoldButterfly}.

The drawback of this concept is in strong limitation on thickness of both the superconducting and ferromagnetic layers and the smallness of the critical currents. 
The thickness of the s layer should be close to its critical value with an accuracy of the order of $ 0.1 \xi \approx 1$  nm, while the thickness of the ferromagnet should ensure the occurence 
of that $0 - \pi$ transition. 
On the other hand, recent experiments \cite{Baek2018, Frolov2018} have demonstrated the feasibility of this task.

The advantages of the proposed SIsFS element compared to spin-valves \cite{Bell, Qader, Baek,  Birge2018} or single flux quantum (SFQ) devices \cite{Krasnov, Averin} are obvious. 
The proposed control memory element stores an information only in the phase difference across the junction in the steady state. Thus, it makes use of its own intrinsic properties to store an information, 
so that there is neither need in remagnetization of the F layers, nor in holding a flux quantum.

In this sense, the bistable  SIsFS device may serve as a truly Josephson memory device, which permits the reduction of size of the auxiliary circuits or even provide an alternative to the SFQ concept of superconducting electronics.

\textbf{Acknowledgments.} The authors acknowledge helpful discussion with V.V. Ryazanov and V.V. Bol'ginov. The design of the memory concept was supported by the Russian Science Foundation (12-17-01079) and numerical study of sFS junction was done with support of RFBR (18-32-00672 mol-a).

\end{document}